\documentclass[twocolumn,showpacs,preprintnumbers,amsmath,amssymb]{revtex4}

\usepackage{graphicx}
\usepackage{dcolumn}
\usepackage{bm}

\begin{document}

\title{Empirical study on clique-degree distribution of networks}
\author{Wei-Ke Xiao$^{1,3}$}
\author{Jie Ren$^{2,3,4}$}
\author{Feng Qi$^3$}
\author{Zhi-Wei Song$^3$}
\author{Meng-Xiao Zhu$^3$}
\author{Hong-Feng Yang$^3$}
\author{Hui-Yu Jin$^3$}
\author{Bing-Hong Wang$^4$}
\author{Tao Zhou$^{2,4}$}
\email{zhutou@ustc.edu}

\affiliation{%
$^1$ Center for Astrophysics, University of Science and Technology
of China, Hefei 230026, PR China \\ $^2$Department of Physics,
University of Fribourg,
Chemin du Muse 3, CH-1700 Fribourg, Switzerland \\
$^3$Research Group of Complex Systems, University of Science and
Technology of China \\ $^4$Department of Modern Physics and
Nonlinear Science Center, University of Science and Technology of
China
}%

\date{\today}

\begin{abstract}
The community structure and motif-modular-network hierarchy are of
great importance for understanding the relationship between
structures and functions. In this paper, we investigate the
distribution of clique-degree, which is an extension of degree and
can be used to measure the density of cliques in networks. The
empirical studies indicate the extensive existence of power-law
clique-degree distributions in various real networks, and the
power-law exponent decreases with the increasing of clique size.
\end{abstract}

\pacs{89.75.Hc, 64.60.Ak, 84.35.+i, 05.40.-a}

\maketitle

The discovery of small-world effect \cite{WS} and scale-free
property \cite{BA} triggered off an upsurge in studying the
structures and functions of real-life networks
\cite{Review1,Review2,Review3,Review4,Review5}. Previous empirical
studies have demonstrated that most real-life networks are
small-world \cite{Amaral2000}, that is to say, it has very small
average distance like completely random networks and large
clustering coefficient like regular networks. Another important
characteristic in real-life networks is the power-law degree
distribution, that is $p(k)\propto k^{-\gamma}$, where $k$ is the
degree and $p(k)$ is the probability density function for the degree
distribution. Recently, empirical studies reveal that many real-life
networks, especially the biological networks, are densely made up of
some functional motifs \cite{Milo2002,BO,Itzkovitz2003}. The
distributing pattern of these motifs can reflect the overall
structural properties thus can be used to classify networks
\cite{Milo2004}. In addition, the networks' functions are highly
affetced by these motifs \cite{Vazquez2004}. A simple measure can be
obtained by comparing the density of motifs between real networks
and completely random ones \cite{Milo2004}, however, this method is
too rough thus still under debate now \cite{Comment,Reply}. In this
paper, we investigate the distribution of \emph{clique-degree},
which is an extension of degree and can be used to measure the
density of cliques in networks.

\begin{table}
\renewcommand{\arraystretch}{1.3}
\caption{The basic topological properties of the present seven
networks, where $N$, $M$, $L$ and $C$ represent the total number
of nodes, the total number of edges, the average distance, and the
clustering coefficient, respectively.} \label{1}
\begin{tabular}{ccccc}
networks/measures & $N$ & $M$ & $L$ & $C$ \\
\hline
Internet at AS level & 10515 & 21455 & 3.66151 & 0.446078 \\
\hline
Internet at routers level & 228263 & 320149 & 9.51448 & 0.060435\\
\hline
the metabolic network & 1006 & 2957 & 3.21926 & 0.216414 \\
\hline
the world-wide web & 325729 & 1090108 & 7.17307 & 0.466293\\
\hline
the collaboration network & 6855 & 11295 & 4.87556 & 0.389773\\
\hline
the ppi-yeast networks & 4873 & 17186 & 4.14233 & 0.122989\\
\hline
the friendship networks & 10692 & 48682 & 4.48138 & 0.178442\\
\hline
\end{tabular}
\end{table}

\begin{figure}
[h]\scalebox{0.3}[0.3]{\includegraphics{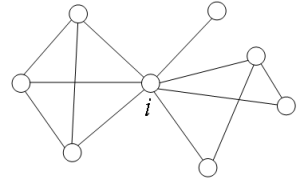}}
\caption{\label{example} Illustration of the clique-degree of node
$i$. $k_i^{(2)}=7$, $k_i^{(3)}=5$, $k_i^{(4)}=1$, and $k_i^{(5)}=0$.
}
\end{figure}

The word \emph{clique} in network science equals the term
\emph{complete subgraph} in graph theory \cite{Derenyi2005}, that is
to say, the $m$ order clique ($m$-clique for short) means a fully
connected network with $m$ nodes and $m(m-1)/2$ edges. Define the
$m$-clique degree of a node $i$ as the number of different
$m$-cliques containing $i$, denoted by $k_i^{(m)}$. Clearly,
2-clique is an edge, and $k_i^{(2)}$ equals to the degree $k_i$,
thus the concept of clique-degree can be considered as an extension
of degree (see Fig. 1). We have calculated the clique-degree from
order 2 to 5 for some representative networks. Figs. 2 to 8 show the
clique-degree distributions of 7 representative networks in
logarithmic binning plots \cite{Park2003,Newman2005}, these are the
Internet at \emph{Autonomous Systems} (AS) level \cite{Data1}, the
Internet at routers level \cite{Data2}, the metabolic network of
\emph{P.aeruginosa} \cite{Data3}, the World-Wide-Web \cite{Data4},
the collaboration network of mathematicians \cite{Data5}, the
protein-protein interaction networks of yeast \cite{Data6}, and the
BBS friendship networks in University of Science and Technology of
China (USTC) \cite{Data7}. The slopes shown in those figures are
obtained by using the maximum likelihood estimation
\cite{Goldstein2004}. Tab. I summarizes the basic topological
properties of those networks.

\begin{figure}
[t]\scalebox{0.7}[0.7]{\includegraphics{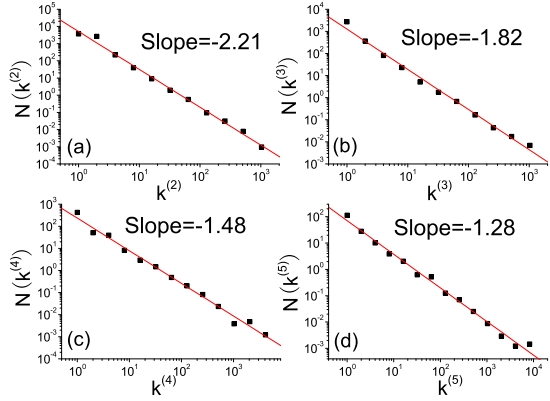}} \caption{(color
online) Clique-degree distributions of Internet at AS level from
order 2 to 5, where $k^{(m)}$ denotes the $m$-clique-degree and
$N(k^{(m)})$ is the number of nodes with m-clique-degree $k^{(m)}$.
In each panel, the marked slope of red line is obtained by using
maximum likelihood estimation \cite{Goldstein2004}.}
\end{figure}

\begin{figure}
[t]\scalebox{0.7}[0.7]{\includegraphics{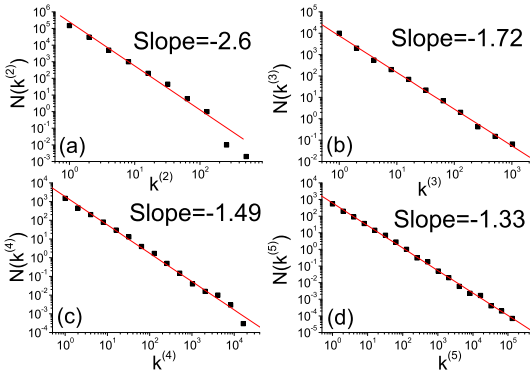}} \caption{(color
online) Clique-degree distributions of Internet at routers level.}
\end{figure}

\begin{figure}
\scalebox{0.7}[0.7]{\includegraphics{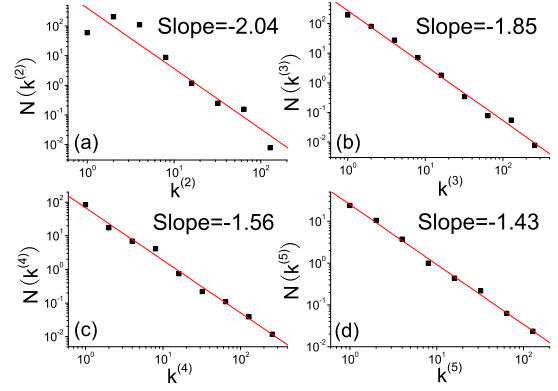}} \caption{(color
online) Clique-degree distributions of the metabolic network of
\emph{P.aeruginosa}}
\end{figure}

\begin{figure}
\scalebox{0.7}[0.7]{\includegraphics{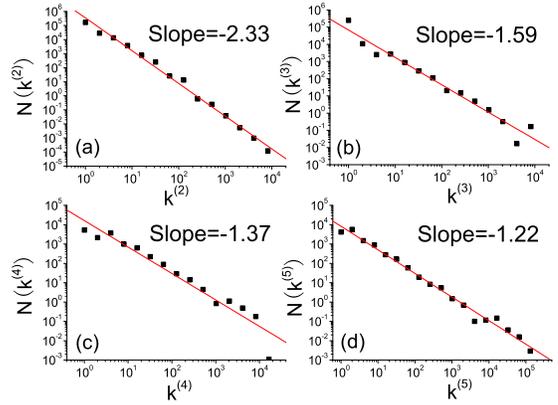}} \caption{(color
online) Clique-degree distributions of the World-Wide Web.}
\end{figure}

\begin{figure}
\scalebox{0.7}[0.7]{\includegraphics{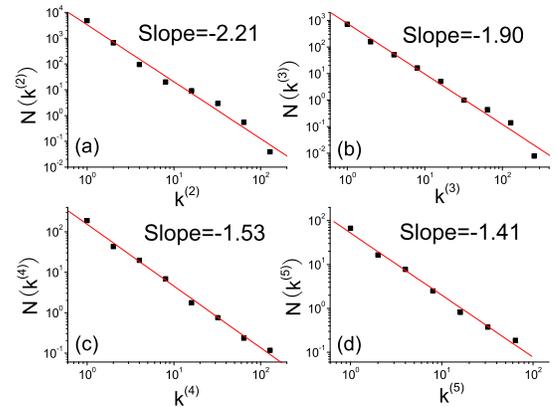}} \caption{(color
online) Clique-degree distributions of the collaboration network
of mathematicians.}
\end{figure}

\begin{figure}
\scalebox{0.7}[0.7]{\includegraphics{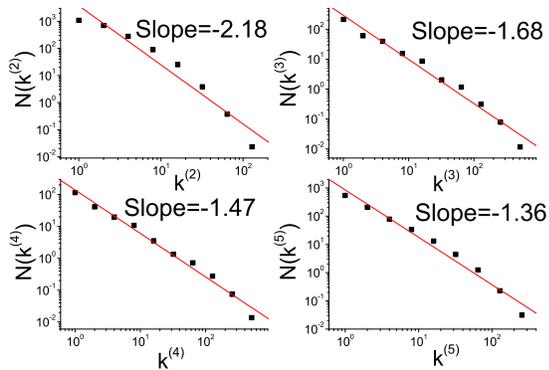}} \caption{(color
online) Clique-degree distributions of the protein-protein
interaction networks of yeast.}
\end{figure}

\begin{figure}
\scalebox{0.7}[0.7]{\includegraphics{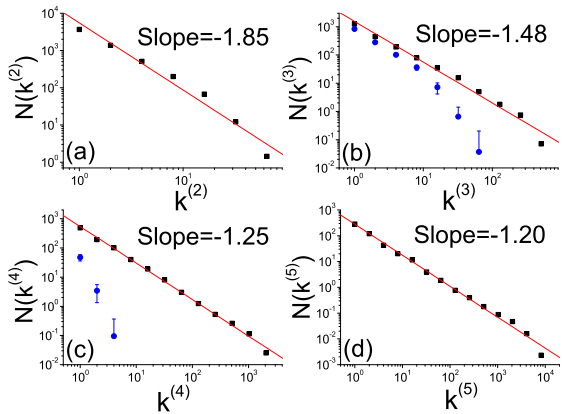}} \caption{(color
online) Clique-degree distributions of the BBS friendship networks
in University of Science and Technology of China. The blue points
with error bars denotes the case of randomized network.}
\end{figure}

Although the backgrounds of those networks are completely different,
they all display power-law clique-degree distributions. We have
checked many examples (not shown here) and observed similar
power-law clique-degree distributions. However, not all the networks
can display higher order power-law clique-degree distributions.
Actually, only the relative large networks could have power-law
clique-degree distribution with order higher than 2. For example,
Ref. \cite{Data3} reports 43 different metabolic networks, but most
of them are very small ($N<1000$), in which the cliques with order
higher than 3 are exiguous. Only the five networks with most nodes
display relatively obvious power-law clique-degree distributions,
and the case of \emph{P.aeruginosa} is shown in Fig. 4. Note that,
even for small-size networks, the high-order clique is abundant for
some densely connected networks such as technological collaboration
networks \cite{Zhang2006} and food webs \cite{Pimm2002}. However,
since the average degree of majority of metabolic networks is less
than 10, the high-order cliques could not be expected with network
size $N<1000$. Furthermore, all the empirical data show that the
power-law exponent will decrease with the increase of clique order.
This may be a universal property and can reveal some unknown
underlying mechanism in network evolution.

\begin{table}
\renewcommand{\arraystretch}{1.3}
\caption{The empirical ($\delta_m$) and predicted ($\delta'_m$)
power-law exponent of clique-degree distribution, where $\gamma$ and
$\alpha$ denote the power-law exponents of degree distribution and
clustering-degree correlation. The symbol ``/" denotes the cases
with $\alpha(m-2)>2$, leading to negative and meaningless
$\delta'_m$. } \label{2}
\begin{tabular}{ccccccc}
networks & $\gamma$ & $\alpha$ & $m$ & $\delta_m$ & $\delta'_m$ & TYPE\\
\hline
Internet at AS level & 2.21 & 1.04 & 3 & 1.82 & 2.26 & II \\
 & & & 4 & 1.48 & / & II \\
 & & & 5 & 1.28 & / & II \\
\hline
Internet at routers level & 2.60 & 0.16 & 3 & 1.72 & 1.86 & I \\
 & & & 4 & 1.49 & 1.63 & I \\
 & & & 5 & 1.33 & 1.53 & I \\
\hline
the metabolic network & 2.04 & 0.80 & 3 & 1.85 & 1.87 & I \\
 & & & 4 & 1.56 & 2.73 & II \\
 & & & 5 & 1.43 & / & II \\
\hline the world-wide web & 2.33 & 1.15 & 3 & 1.59 & 2.56 & II \\
 & & & 4 & 1.37 & / & II \\
 & & & 5 & 1.22 & / & II \\
\hline
the collaboration network & 2.21 & 0.90 & 3 & 1.90 & 2.10 & II \\
 & & & 4 & 1.53 & 5.03 & II \\
 & & & 5 & 1.41 & / & II \\
\hline
the ppi-yeast networks & 2.18 & 0.91 & 3 & 1.68 & 2.08 & II \\
 & & & 4 & 1.47 & 5.37 & II \\
 & & & 5 & 1.36 & / & II \\
\hline
the friendship networks & 1.85 & 0.32 & 3 & 1.48 & 1.51 & I \\
 & & & 4 & 1.25 & 1.42 & I \\
 & & & 5 & 1.20 & 1.41 & I \\
\hline
\end{tabular}
\end{table}

In order to illuminate that the power-law clique-degree
distributions with order higher than 2 could not be considered as a
trivial inference of the scale-free property, we compare these
distributions between original USTC BBS friendship network and the
corresponding randomized network. Here the randomizing process is
implemented by using the edge-crossing algorithm
\cite{Milo2004,Reshuffle1,Reshuffle2,Zhao2006}, which can keep the
degree of each node unchanged. The procedure is as follows: (1)
Randomly pick two existing edges $e_1=x_1x_2$ and $e_2=x_3x_4$, such
that $x_1\neq x_2\neq x_3\neq x_4$ and there is no edge between
$x_1$ and $x_4$ as well as $x_2$ and $x_3$. (2) Interchange these
two edges, that is, connect $x_1$ and $x_4$ as well as $x_2$ and
$x_3$, and remove the edges $e_1$ and $e_2$. (3) Repeat (1) and (2)
for $10M$ times.

We call the network after this operation the \emph{randomized
network}. In Fig. 9, we report the clique-degree distributions in
the randomized network. Obviously, the 2-clique degree distribution
(not shown) is the same as that in Fig. 8. One can find that the
randomized network does not display power-law clique-degree
distributions with higher order, in fact, it has very few 4-cliques
and none 5-cliques. The direct comparison is shown in Fig. 8.

\begin{figure}
\scalebox{0.7}[0.7]{\includegraphics{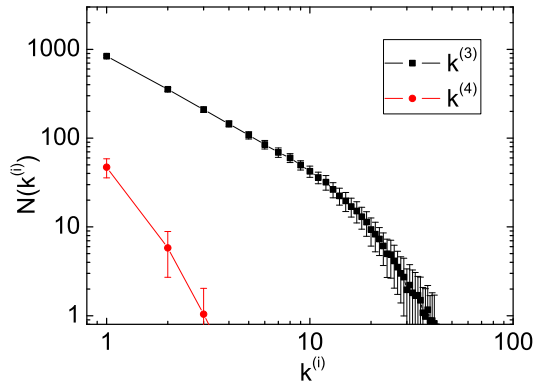}} \caption{(color
online) The clique-degree distributions in the randomized network
corresponding to the BBS friendship network of USTC. The black
squares and red circles represent the clique-degree distributions of
order 3 and 4, respectively. All the data points and error bars are
obtained from 100 independent realizations.}
\end{figure}

The discoveries of new topological properties of networks infuse the
network science with ozone
\cite{WS,BA,Review5,Milo2002,Newman2002,Ravasz2003,Song2005}. These
empirical studies not only reveal new statistical features of
networks, but also provide useful criterions in judging the validity
of evolution models (For example, the Barab\'{a}si-Albert model
\cite{BA} does not display high order power-law clique-degree
distributions.). The clique-degree, which can be considered as an
extension of degree, may be useful in measuring the density of
motifs, such subunits not only plays a role in controlling the
dynamic behaviors, but also refers the basic evolutionary
characteristics. More interesting, we find various real-life
networks display power-law clique-degree distributions of decreasing
exponent with the clique order. This is an interesting statistical
property, and can provide a criterion in the studies of modelling.

It is worthwhile to remind of a prior work \cite{Vazquez2004} that
reported a similar power-law distribution observed for some cellular
networks. They divided all the subgraphs into two types, and claim
that the power-law can only be found in TYPE I. Moreover, they have
derived the analytical expression of the power-law exponent
$\delta'_m$ for $m$-clique degree distribution as \cite{Vazquez2004}
$\delta'_m=1+(\gamma-1)/[m-1-\alpha (m-1)(m-2)/2]$, where $\alpha$
denotes the power-law exponent of clustering-degree correlation
$C(k)\sim k^{-\alpha}$. Tab. II displays the predicted power-law
exponents $\delta'_m$, compared with the empirical observation
$\delta_m$. For the TYPE I cases, the predicted results are, to some
extent, in accordance with the empirical data. More significant,
here we offer an clear evidence that those power-laws can also be
detected for TYPE II subgraphs, while Ref. \cite{Vazquez2004}
claimed that the power law can not be observed for TYPE II cases.
Note that, even the power law is detected for TYPE II cases, the
analytical expression of $\delta'_m$ loses its validity in those
cases. The qualitative difference in TYPE II cases and quantitative
departure in TYPE I cases may be attributable to the structural bias
(e.g. assortative connecting pattern \cite{Newman2002}, rich-club
phenomenon \cite{Zhou2004}, etc.) since the derivation in Ref.
\cite{Vazquez2004} is based on uncorrelated networks. In addition,
the predicted accuracy decreases as the increase of clique size $m$,
because the clustering coefficient takes into account only the
triangles \cite{ex}. Therefore, a more accurate analysis may involve
higher order clustering coefficient \cite{Review5}. In a word, Ref.
\cite{Vazquez2004} provides us a start point of in-depth
understanding on network structure in clique level, while the
diversity and complexity of real networks require further
explorations on this issue.

We thank Dr. Ming Zhao for the useful discussion. This work is
supported by the National Natural Science Foundation of China under
Nos. 10472116, 70471033, and 10635040.


\begin{thebibliography}{1}
\bibitem{Data7} This network is constructed based on the BBS of USTC, wherein each node represents a BBS accounts and two nodes are neighboring if one appears in the other one's friend-list. Only the undirected network is considered.
\end{thebibliography}
\end{document}